\documentstyle[prc,aps,preprint]{revtex}
\begin{document}
\draft
\date{\today }
\title{On the Collisional Damping of Giant Dipole Resonance}

\author{Osman Yilmaz$^{1}$, Ahmet Gokalp$^{1}$, Serbulent Yildirim$^{1}$ and Sakir Ayik$^{2}$}
\address{{\small $^{1}$ {\it Physics Department, Middle East Technical University,}\\
06531 Ankara, Turkey}}
\address{{\small $^{2}$ {\it Physics Department, Tennessee Technological University,}
Cookeville TN 38505, USA}}
\maketitle

\begin{abstract}
Collisional damping widths of giant dipole excitations are calculated in
Thomas-Fermi approximation by employing the microscopic in-medium 
cross-sections of Li and Machleidt and the phenomenological Gogny force. The results 
obtained in both calculations compare well, but account for about $25-35\%$ 
of the observed widths in $^{120}Sn$ and $^{208}Pb$ at finite temperatures.
\end{abstract}

\pacs{ 21.30.Fe, 24.30.Cz}

 
During last several years, much work has been done to
understand the damping properties of giant resonance excitations at zero and
finite temperature \cite{Hofmann,TinLead}. In medium-weight and heavy nuclei
at relatively low temperatures the overwhelming contribution to damping
arises from the spreading width $\Gamma^{\downarrow}$ due to mixing of
collective excitations with more complicated doorway states. There are
essentially two different approaches for calculation of the spreading widths: 
(i) Coherent mechanism due to coupling with low-lying
surface modes which provides an important mechanism for damping of giant
resonance in particular at low temperatures \cite
{BerBorBro,Wambach,Lauritzen,Chomaz}, (ii) Damping due to the coupling with
incoherent 2p-2h states which is usually referred to as the collisional
damping \cite{BelAyi,AyiYil,Mora,LacCho}, and the Landau damping modified
by two-body collisions \cite{DiToro,Baran,Shlomo}. Calculations carried out
on the basis of these approaches are partially successful in explaining the
broadening of the giant dipole resonance (GDR) with increasing excitation energy.
In particular, the model based on the coupling with thermal shape
fluctuations in adiabatic approximation provides a reasonable description
for the dipole strength distribution in $^{208} Pb$ and the $^{120}Sn$
nuclei \cite{Ormand1,Ormand2,Dimitri}. In this work, we do not consider the
coherent contribution to damping, but investigate the collisional damping at
finite temperature due to decay of the collective state into incoherent
2p-2h excitations in the basis of a non-Markovian transport approach. In
order to assess how much of the total width of giant resonance excitations
is exhausted by decay into the incoherent 2p-2h states, we need 
realistic in-medium cross-sections which interpolate correctly between the
free space and the medium. In previous investigations, incoherent
contributions have been estimated by employing either free nucleon-nucleon
cross sections or an effective Skyrme force, 
which provide, at most, a semi-quantitative description of the 
collisional damping widths \cite{AyiLac2}. In this respect, the Skyrme force provides a poor approximation
in the collision term, since in the vicinity of nuclear surface, it does not
match at all to the free nucleon-nucleon cross-sections. In this letter, we
present calculations performed in Thomas-Fermi approximation by employing
the microscopic in-medium cross-sections calculated by Li and Machleidt \cite
{Li}, which interpolate correctly between the free space and the medium, and
provide the best available input for determining the magnitude of the
collisional damping at finite temperature. For comparison, we also carry out
calculations using the phenomenological finite range Gogny force and the
zero range Skyrme force \cite{Schuck,Ring}.

The small amplitude limit of the extended TDHF provides an appropriate basis
for investigating collective vibrations, in which damping due the incoherent
2p-2h decay is included in the form of a non-Markovian collision term \cite
{Abe}. Then, a description for small density fluctuations, $\delta
\rho(t)=\rho (t)-\rho _{0}$, is obtained by linearizing the extended TDHF
theory around a finite temperature equilibrium state $\rho _{0}$ \cite
{AyiYil,AyiLac1}, 
\begin{eqnarray}
\lefteqn{i\hbar \frac{\partial }{\partial t}\delta \rho -[h_{0},\delta \rho
]- [\delta U,\rho _{0}] = } \\
&& -\frac{i}{\hbar }\int^{t}dt^{\prime}Tr_{2}\left[v, \rho _{1}^{0}\rho
_{2}^{0}e^{-ih_{0}(t-t^{\prime })}[Q(t^{\prime }),v]e^{+ih_{0}(t-t^{\prime
})}(1-\rho _{1}^{0})(1-\rho _{2}^{0})\right] - h.c.  \nonumber
\end{eqnarray}
where $\delta U $ represents small deviations in the effective mean-field
potential. The right hand side describes a non-Markovian quantal collision
term in which the quantity $Q(t)$ denotes the distortion matrix associated
with the single particle density matrix, $\delta \rho (t)\equiv [Q(t),\rho
_{0}]$. It is convenient to analyze collective vibrations by expanding
the small deviation $\delta \rho (t)$ in terms of finite temperature RPA
modes $O_{\lambda }^{\dagger }$ and $O_{\lambda }$, 
\begin{equation}
\delta \rho (t)=\sum_{\lambda >0}z_{\lambda }(t)[O_{\lambda }^{\dagger
},\rho _{0}]-z_{\lambda }^{*}(t)[O_{\lambda },\rho _{0}]
\end{equation}
where $z_{\lambda }(t)$ and $z_{\lambda }^{*}(t)$ denote the amplitudes
associated with the RPA modes, $O_{\lambda }^{\dagger }$ and $O_{\lambda }$,
which are determined by the finite temperature RPA equation \cite
{Vaut,Sagawa}. Substituting this expansion into eq.(1) and projecting by $%
O_{\lambda }$, and noting that $Q(t)=\sum z_{\lambda }(t)O_{\lambda
}^{\dagger }-z_{\lambda }^{*}(t)O_{\lambda }$, we find that the amplitudes
of the RPA modes execute damped harmonic motion \cite{AyiYil,AyiLac2} 
\begin{equation}
i\hbar \frac{d}{dt}z_{\lambda }-\hbar \omega _{\lambda }z_{\lambda
}=\int^{t}dt^{\prime }\Sigma _{\lambda }(t-t^{\prime })z_{\lambda
}(t^{\prime })
\end{equation}
where the right hand side describes the self-energy of the collective mode
due to coupling to 2p-2h excitations. In the Hartree-Fock representation,
the Fourier transform of the self-energy is given by, 
\begin{equation}
\Sigma _{\lambda }(\omega )=\frac{1}{4}\sum \frac{\left|<ij|[O_{\lambda
}^{\dagger },v]|kl>_{A}\right|^{2}}{\hbar \omega -\Delta \epsilon +i\eta }[
n_{k}n_{l}\bar{n}_{i}\bar{n}_{j}-n_{i}n_{j}\bar{n}_{k}\bar{n}_{l}]
\end{equation}
where $\bar{n}_{i}=1-n_{i}$, $\Delta \epsilon =\epsilon _{i}+\epsilon
_{j}-\epsilon_{k}-\epsilon _{l}$ , and $\eta $ is a small positive number.
The self-energy agrees with  the expression written down by Landau
\cite{Landau}, in which the real and imaginary parts, $\Sigma
_{\lambda }(\omega )= \Delta _{\lambda }(\omega )-\frac{i}{2}\Gamma_{\lambda
}(\omega )$, determine the energy shift and the damping width of the
collective excitation, respectively \cite{Wambach}.

We consider the expression for the self-energy in the Thomas-Fermi
approximation, which corresponds to a semi-classical transport description
of the collective vibrations. It can be shown that the 2p-2h self-energy in
Thomas-Fermi approximation can be deduced from the quantal expression by
replacing the occupation numbers with the equilibrium phase-space 
density given by the Fermi-Dirac function $n_i \rightarrow 
f(\epsilon, T)=1/[exp(\epsilon-\mu)/T+1]$  with $\mu$ as the chemical potential, and summations over the 2p-2h states with
integrals over phase-space, $\Sigma \rightarrow \int d^3r d^3p_1 d^3p_2
d^3p_3 d^3p_4 $ \cite{AyiYil,Hasse}. Furthermore, spin-isospin effects can
be included into the treatment by considering proton and neutron degrees of
freedom separately. The small deviations of the density matrices $\delta
\rho_{p}(t)$, $\delta \rho_{n}(t)$ of protons and neutrons are determined by
two coupled equations analogous to eq.(1). The collision terms in these
equations involve binary collisions between proton-proton, neutron-neutron
and proton-neutron, and a summation over the spins of the colliding
particles. Since in isoscalar/isovector modes, proton and neutron densities
vibrate in-phase/out-of phase, $\delta \rho_{p}(t)=\mp \delta \rho_{n}(t)$,
we can deduce equation of motions for describing isoscalar/isovector
vibrations by adding and subtracting the corresponding equations for protons
and neutrons. As a result, in Thomas-Fermi approximation, we obtain for the
collisional widths of isovector modes \cite{AyiYil},
\begin{equation}
\Gamma_{\lambda}^{v}= \frac {1}{N_{\lambda}} \int d{\bf p}_{1} d{\bf p}_{2} d%
{\bf p}_{3} d{\bf p}_{4} [(W_{pp}+W_{nn}) \left ( \frac {\Delta
\chi_{\lambda}}{2}\right )^{2}+ 2W_{pn}\left ( \frac {\widetilde {\Delta \chi%
}_{\lambda}}{2} \right )^{2}] Z f_{1}f_{2}\bar{f}_{3}\bar{f}_{4}
\end{equation}
and a similar expression for the isoscalar modes. Here, $N_{\lambda}=\int d%
{\bf r} d{\bf p}(\chi_{\lambda})^{2} (-\frac{\partial}{\partial \epsilon}f)$
is a normalization, $\Delta
\chi_{\lambda}=\chi_{\lambda}(1)+\chi_{\lambda}(2)-
\chi_{\lambda}(3)-\chi_{\lambda}(4)$, $\widetilde{\Delta \chi}%
_{\lambda}=\chi_{\lambda}(1)-\chi_{\lambda}(2)- \chi_{\lambda}(3)+
\chi_{\lambda }(4)$, $Z= [\delta(\hbar\omega_{\lambda}-\Delta\epsilon)-
\delta(\hbar\omega_{\lambda}+\Delta\epsilon)]/ \hbar\omega_{\lambda}$, and $%
\chi_{\lambda}(t)$ denotes the distortion factor of the phase-space density $%
\delta f(t)= \chi_{\lambda}(t) (-\partial f/\partial \epsilon)$ in the
corresponding mode. In this expression, transition rates $W_{pp}, W_{nn},
W_{pn}$ associated with proton-proton, neutron-neutron and proton-neutron
collisions are given in terms of the corresponding scattering cross-sections as,
\begin{equation}
W(12;34)= \frac{1}{(2\pi\hbar)^3}\frac{4\hbar}{m^2}\frac{d\sigma}{d\Omega}
\delta({\bf p}_{1}+{\bf p}_{2}-{\bf p}_{3}-{\bf p}_{4}).
\end{equation}

We apply the formula (5) to calculate the collisional widths of  the GDR 
excitations by taking distortion factors of the momentum distribution
according to the scaling picture as $\chi_{D}({\bf p})={\bf p \cdot e}$,
where ${\bf e}$ denotes the unit vector along the relative displacement of the proton
and neutron momentum distributions. 
Due to momentum conservation, terms involving $W_{pp}$ and $W_{nn}$ drop
out, and the damping is determined by the proton-neutron collision term. The
spin averaged proton-neutron cross-section associated with an effective
residual interaction can be expressed as, 
\begin{equation}
\left(\frac{d\sigma}{d\Omega}\right)_{pn}= \frac{\pi}{(2\pi\hbar)^3}\frac{m^2%
}{4\hbar}\cdot \frac{1}{8}\sum_{S,T} (2S+1)|<{\bf q};S,T|v |{\bf q}%
^{\prime};S,T>_{A}|^{2}
\end{equation}
where ${\bf q}=({\bf p}_{1}-{\bf p}_{2})/2$, ${\bf q}^{\prime}=({\bf p}_{3}-%
{\bf p}_{4})/2$ are the relative momenta before and after a binary
collision, and $<{\bf q};S,T|v|{\bf q}^{\prime};S,T>_{A}$ represents the
fully anti-symmetric matrix element of the residual interaction between two
particle states with total spin and isospin $S$ and $T$. 
 By noting that, S=T=1 and S=T=0 matrix elements of the interaction are space
anti-symmetric, and S=1, T=0 and S=0, T=1 matrix elements are space
symmetric, we find that proton-neutron cross-section associated with the
Gogny force is given, 
\begin{eqnarray}
\left( \frac{d\sigma}{d\Omega}\right)_{pn}^G & = & \frac{\pi}{(2\pi\hbar)^3}%
\frac{{m_G^*}^2}{4\hbar}\frac{1}{8} \left \{ 3 \left| \sum_{i=1}^2
I_i^{-}(W_i + B_i - H_i - M_i ) \right|^2+ \left| \sum_{i=1}^2 I_i^{-}(W_i -
B_i + H_i -M_i ) \right|^2+ \right.  \nonumber \\
& & \left. 3 \left| \sum_{i=1}^2 I_i^{+}( W_i + B_i + H_i + M_i) +4 t_3
\rho^{1/3}\right|^2+ \left| \sum_{i=1}^2 I_i^{+}(W_i - B_i - H_i + M_i
)\right|^2 \right \}
\end{eqnarray}
where $m_G^{*}$ denotes the effective mass corresponding to the Gogny force,
and the quantities $I_i^{+}$ and $I_i^{-}$ are the symmetric and
anti-symmetric matrix elements of the Gaussian factor in the force, 
\begin{equation}
I_{i}^{\pm}= (\sqrt{\pi}\mu_i)^3 \left( exp [- \frac{1}{4} ({\bf q}-{\bf q}%
^{\prime})^2 (\frac {\mu_i}{\hbar})^2] \pm exp [- \frac{1}{4} ({\bf q}+ {\bf %
q}^{\prime})^2 (\frac {\mu_i}{\hbar})^2] \right).
\end{equation}
In these expressions, $\rho$ is the local density and $W_i , B_i , H_i ,
M_i, \mu_i $ denote the standart parameters of the Gogny force \cite{Schuck,Ring}. 
In a similar manner, we can calculate the cross-section associated with the
Skyrme force and obtain \cite{AyiYil}, 
\begin{eqnarray}
\left( \frac{d\sigma}{d\Omega}\right)_{pn}^S = \frac{\pi}{(2\pi\hbar)^3} 
\frac{{m_S^*}^2}{4\hbar}\frac{1}{2} & &\left( [t_0 (1-x_0 )+\frac{t_1}{%
2\hbar^2} (1-x_1)({\bf q}^2+{{\bf q}^\prime}^2)+ \frac{t_3}{6}%
(1-x_3)\rho^{\alpha}]^2+ \right. \\
& & \left. 3[t_0 (1+x_0 )+ \frac{t_1}{2\hbar^2}(1+x_1)({\bf q}^2+{{\bf q}%
^\prime}^2)+ \frac{t_3}{6}(1+x_3) \rho^{\alpha}]^2 + \right.  \nonumber \\
& & \left. [ (\frac {t_2}{\hbar^2})^2 (1-x_2)^2 + 3 (\frac {t_2}{\hbar^2})^2
(1+x_2)^2] ({\bf q} \cdot {\bf q}^\prime)^2 \right)  \nonumber
\end{eqnarray}
The expressions for the effective masses $m_G^*( r )$ and $m_S^*( r )$ for
the Gogny and Skyrme forces can be found in \cite{Schuck,Ring}. 
The angle $\Theta$ between ${\bf q}$ and ${\bf q}^{\prime}$ in the
cross-sections (9) and (13) defines the scattering angle in the center of
mass frame, and the total cross-section is found by an integration over this
angle, $\sigma_{pn}=2\pi \int sin\Theta d\Theta \left( d\sigma/d\Omega
\right)_{pn}$. This integral is trivial for the Skyrme cross-section, but a
numerical integration is needed over the scattering angle to find the total
proton-neutron cross-section for the Gogny force.

In figures 1 and 2, we compare the microscopic in-medium cross-sections of Li
and Machleidt with the cross-sections (dotted lines) of the Gogny force
(dashed lines) and the Skyrme force (solid lines) with the SkM$^*$
parameters. For this purpose we employ a convenient parameterization of the
microscopic calculations as reported in \cite{Li}, 
\begin{equation}
\sigma_{pn}^{LM}= [31.5+0.092 \left|20.2-E_{lab}^{0.53}\right|^{2.9}] \; 
\frac{1.0+0.0034 \; E_{lab}^{1.51}\; \rho^2} {1.0+21.55  \; \rho^{1.34}}
\end{equation}
where $E_{lab}=({\bf p}_1-{\bf p}_2)^2/2m=2 {\bf q}^2/m$ is the kinetic
energy of the projectile in the rest frame of the target nucleon which is
also equal to twice the energy available in the center of mass frame. Figure
1 illustrates the total cross-sections as a function of the bombarding
energy $E_{lab}$ at two different nuclear matter densities $%
\rho=\rho_{0}=0.18 fm^{-3}$ (top panel) and $\rho=\rho_{0}/2$ (bottom
panel). The cross-sections shown in the left and right parts of the figure
are calculated with the bare nucleon mass and the corresponding effective
masses, respectively. Around the normal matter density $\rho \approx \rho_{0}$, 
and over a narrow energy interval around 150 MeV, these cross-sections roughly match, however the phenomenological cross-sections deviate strongly from the microscopic 
cross-sections at lower densities and lower and higher energies. Discrepancies are even larger in the calculations with the bare nucleon mass. In figure 2, the cross-sections are plotted as a function of density at the bombarding energy $E_{lab}=100$ MeV. For decreasing density, the microscopic calculations approach the free proton-neutron cross-section and compare well with the experimental data, where as the phenomenological
cross-sections strongly increase and reach large values in free space. From these observations, we can safely state that the microscopic calculations of Li and Macleidt provide a more reliable description of the in-medium cross-sections than those given by the Gogny and Skyrme type forces.

In a previous work, the momentum integrals in the expression (5) for the
damping width are evaluated in an approximate manner using a method familiar
from the Fermi liquid theory. The method subsequently improved by
incorporating the surface corrections \cite{AyiYil}. Here, we perform the momentum
integrals exactly. For this purpose it is convenient to transform the
variable into the relative momentum and total momentum before and after the
collision, ${\bf q}=({\bf p}_1-{\bf p}_2)/2$, ${\bf P}={\bf p}_1+{\bf p}_2$,
and ${\bf q}^{\prime }=({\bf p}_3-{\bf p}_4)/2$, ${\bf P}^{\prime }={\bf p}%
_3+{\bf p}_4$. Then, in expression (5), the integral over ${\bf P}^{\prime }$
is done immediately to give 
\begin{equation}
\Gamma _D=\frac 2{N_D}\int d^3r[\frac 1{(2\pi \hbar )^3}\frac{4\hbar }{{m^{*}%
}^2}]\frac 1{\hbar \omega _D}[1-exp(-\frac{\hbar \omega _D}T)]I_D(r)
\end{equation}
with 
\begin{equation}
I_D(r)=\int d^3Pd^3qd^3q^{\prime }({\bf q\cdot e}-{\bf q^{\prime }\cdot e}%
)^2\left( \frac{d\sigma }{d\Omega }\right) _{pn}\delta (\hbar \omega
_D-\epsilon ^{\prime }+\epsilon )f_1f_2\bar{f}_3\bar{f}_4
\end{equation}
where $\epsilon ={\bf q}^2/m^{*}$ and $\epsilon ^{\prime }={{\bf q}^{\prime }%
}^2/m^{*}$ represent energies of two particle system in the center of mass
frame before and after the collision. In terms of the integration variables,
the products of the phase-space factors have the following forms, 
\begin{equation}
f_1f_2=\frac 1{1+A^2(E,\epsilon )+A(E,\epsilon )\left( exp(+z\sqrt{E\epsilon 
}/T)+exp(-z\sqrt{E\epsilon })/T\right) }
\end{equation}
and 
\begin{equation}
\bar{f}_3\bar{f}_4=\frac{A^2(E,\epsilon ^{\prime })}{1+A^2(E,\epsilon
^{\prime })+A(E,\epsilon ^{\prime })\left( exp(+z^{\prime }\sqrt{E\epsilon
^{\prime }})/T+exp(-z^{\prime }\sqrt{E\epsilon ^{\prime }})/T\right) }
\end{equation}
where $A(E,\epsilon )=exp\left( (E+\epsilon )/2-\epsilon _F\right) /T$, and $%
z=cos\theta $ and $z^{\prime }=cos\theta ^{\prime }$ are the angles between $%
{\bf q}$ and ${\bf P}$, and ${\bf q}^{\prime }$ and ${\bf P}$, respectively.
In the further evaluation of the momentum integrals, we neglect the angular
anisotropy of the cross-sections and make the replacement $(d\sigma /d\Omega
)_{pn}\rightarrow \sigma _{pn}/4\pi $ to obtain, 
\begin{equation}
I_D(r)=(2\pi )^2(m^{*}\sqrt{m^{*}})^3\frac{m^{*}}3\int_0^\infty
\int_0^\infty \sqrt{E}dE\sqrt{\epsilon \epsilon ^{\prime }}d\epsilon \sigma
_{pn}(\epsilon +\epsilon ^{\prime })\int_{-1}^{+1}\int_{-1}^{+1}dzdz^{\prime
}f_1f_2\bar{f}_3\bar{f}_4
\end{equation}
where $\epsilon ^{\prime }=\epsilon +\hbar \omega _D$ and $E=P^2/4m^*$ denotes
the kinetic energy of the center of mass. In obtaining this result,
we make the  replacement 
$({\bf q\cdot e}-{\bf q^{\prime }\cdot e})^2 \rightarrow
({\bf q}^2+{\bf q^{\prime }}^2)/3$ in the integrand of (13), which follow from the fact that the integral of the cross terms vanishes and the integral is independent of the orientation of the unit vector ${\bf e}$. It is possible to carry out the angular integrals over $z$ and $z^{\prime }$ analytically, and we perform the remaining three fold integrations over $E$, $\epsilon $ and the radial coordinate $r$ numerically.

In the numerical evaluations, we determine the nuclear density $\rho( r )$ in Thomas-Fermi approximation using a Wood-Saxon potential with a depth $V_0=-44$ MeV, thickness $a=0.67$ fm and sharp radius $R_0= 1.27 A^{1/3}$ fm, calculate the
position dependent chemical potential $\mu(r,T)$ in the Fermi-Dirac function
$f(\epsilon, T)$ at each temperature, and
use the formula $\hbar \omega=80 A^{-1/3}$ MeV for the GDR energies. In Fermi gas picture, the effective excitation energy of the system and temperature are related according to, 
\begin{equation}
E^*=\int \frac{4}{(2\pi\hbar)^3} d^3r d^3p \; \epsilon \; 
[f(\epsilon, T)-f(\epsilon, T=0)]
\end{equation}
For temperatures small compared to the Fermi energy $\epsilon_F=\mu(T=0)$ 
and in sharp radius
approximation, the relation between the excitation energy and temperature
becomes, $E^* = a_F T^2$, where $a_F= A \pi^2/4\epsilon_F$ is the Fermi gas
level density parameter. The experimental temperature $T^*$ is determined
from the excitation energy using a similar relation $T^*=\sqrt{E^*/a_E}$,
where $a_E$ denotes the energy dependent empirical level density parameter \cite{TinLead}.
Hence, the temperature parameter in $f(\epsilon, T)$ is related to the experimental
temperature as $T=T^*\sqrt{a_E/a_F}$. Figure 3 shows the collisional damping
width of GDR in $^{120}Sn $ and $^{208}Pb $ as a function of the
experimental temperature and comparison with data. Calculations performed with the cross-sections of Li and Machleidt are indicated with dotted lines. For comparison, we also indicate the  results with the SkM$^*$ (solid lines) and the Gogny (dashed lines) cross-sections with the bare nucleon mass (lower panel) and the effective nucleon mass (upper panel), and the points are the experimental data \cite{TinLead}. The calculations with cross-sections of Li and Machleidt exhibit a weaker temperature dependence than data and account for about $25-35\%$ of the experimental damping widths in
tin and lead nuclei at finite temperatures. As discussed above, the calculations with Gogny and Skyrme forces are not reliable, since the phenomenological cross-sections do not interpolate correctly between the free space and the medium. Their magnitude at low densities and in the vicinity of Fermi energy, where the dominant contributions to damping arises, become much larger than the cross-sections of Li and Machleidt. As a result, the phenomenological forces predict larger damping, although, the magnitude of the damping is reduced by the effective mass, as seen from the top panel of figure 3.  However, it is surprising that, the damping width calculated with the Gogny force including the effective mass reduction almost agree with the results obtained with the cross-sections of the Li and Macleidt for both tin and lead nuclei.
Figure 4 illustrates a comparison of the calculations with the revised data taken from \cite{Dimitri}. In the revised data, the effective temperatures are smaller and the
damping widths show a stronger temperature dependence  than reported in \cite{TinLead}. Consequently, the collisional damping accounts for a smaller fraction of  the experimental damping widths.

In order  to assess the fraction of the total width of collective excitations that is exhausted by decay into the incoherent 2p-2h states, namely the collisional damping, we need realistic in-medium nucleon-nucleon cross-sections around Fermi energy. 
In this respect, microscopic in-medium cross-sections of Li and Machleidt provides the best available input, since the cross-sections interpolate correctly between the free space
and the medium. In contrast, the cross-sections based on the phenomenological Skyrme force and even the finite range Gogny force exhibit unrealistic behavior as a function of density and energy.  In the present investigations, by employing cross-sections of Li and Machleidt, we carry out calculations of the damping widths of giant dipole excitations in Thomas-Fermi approximation at finite temperature, and compare the results with the GDR measurements in$^{120}Sn$ and $^{208}Pb$ nuclei. We also compare the results with calculations based on the Skyrme and Gogny cross-sections. To the extend that we accept the validity of the in-medium cross-sections of Li and Machleidt, 
we can conclude that 
the collisional damping of the GDR excitations is not very strong, and accounts for about 1/3 of $\Gamma^{\downarrow}$ in both nuclei at zero and finite temperatures. Consequently, there is plenty of room remains for coherent damping mechanism due to coupling of the dipole mode with surface fluctuations. 

\begin{center}
{\bf Acknowledgments}
\end{center}

One of us (S. A.) gratefully acknowledges the Physics Department of Middle
East Technical University for warm hospitality extended to him during his
visits. We thank D. Kusnezov for providing the revised data of $^{120}Sn$ and $^{208}Pb$, R. Machleidt for providing a table of their cross-sections, and P. Schuck 
for fruitful discussions. This work is supported in part by the U.S. DOE Grant No.
DE-FG05-89ER40530.


\newpage

{\bf Figure Captions:}

\begin{description}
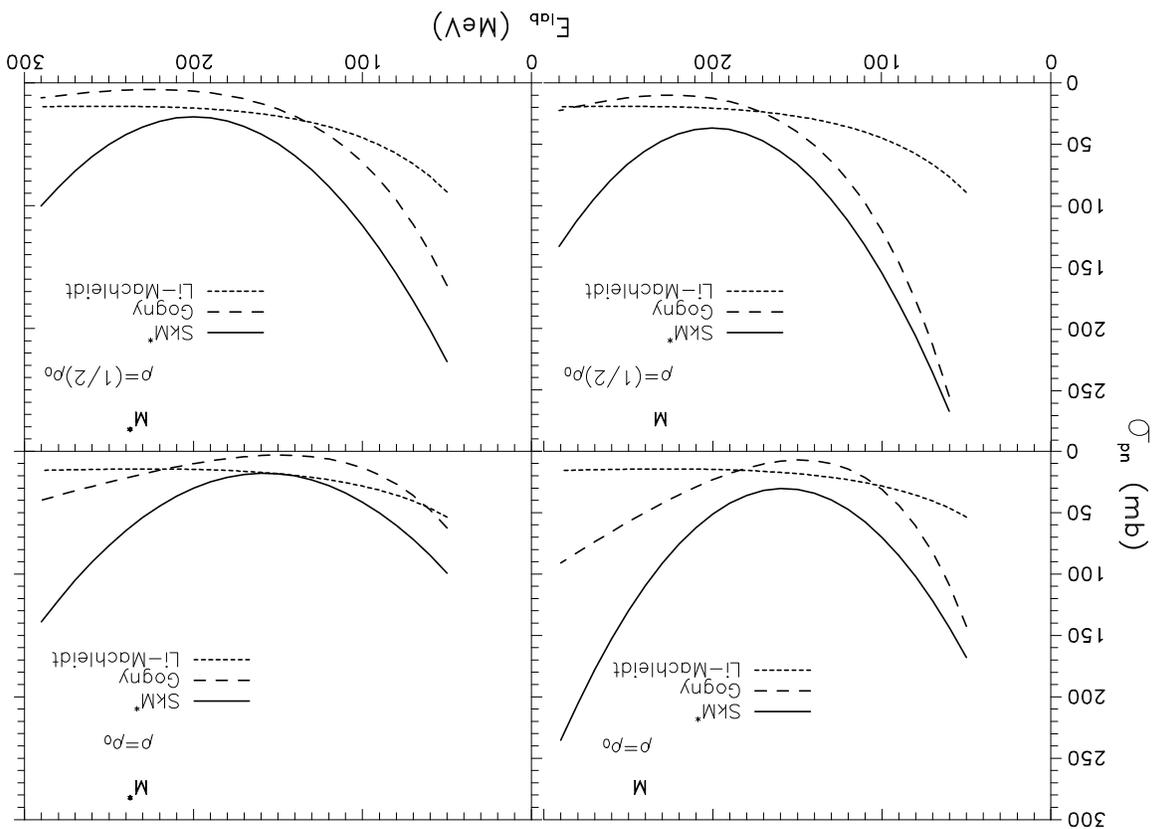
\item[{\bf Figure 1}:]  The proton-neutron in-medium cross-sections as a
function of bombarding energy $E_{lab}$ at several different densities.
Dotted lines are cross-sections of Li and Machleidt, and solid and dashed
lines are cross-sections associated with the SkM$^{*}$ and the Gogny forces
with the bare nucleon mass (left) and the effective nucleon mass (right),
respectively.

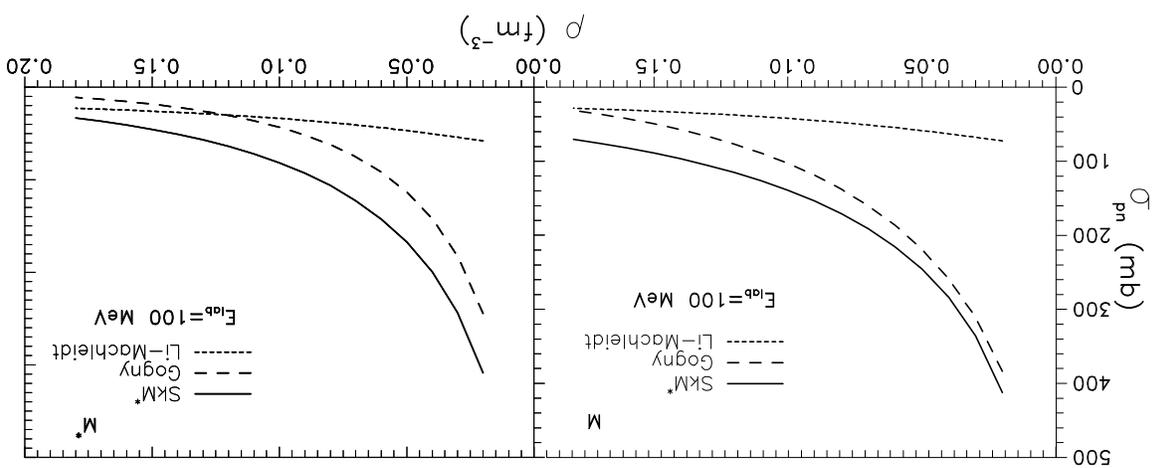
\item[{\bf Figure 2}:]  The proton-neutron in-medium cross-sections as a
function of density $\rho $ at $E_{lab}=100MeV$. Dotted lines are
cross-sections of Li and Machleidt, and solid and dashed lines are
cross-sections with the SkM$^{*}$ and the Gogny forces with the bare nucleon
mass (left) and the effective nucleon mass (right), respectively.

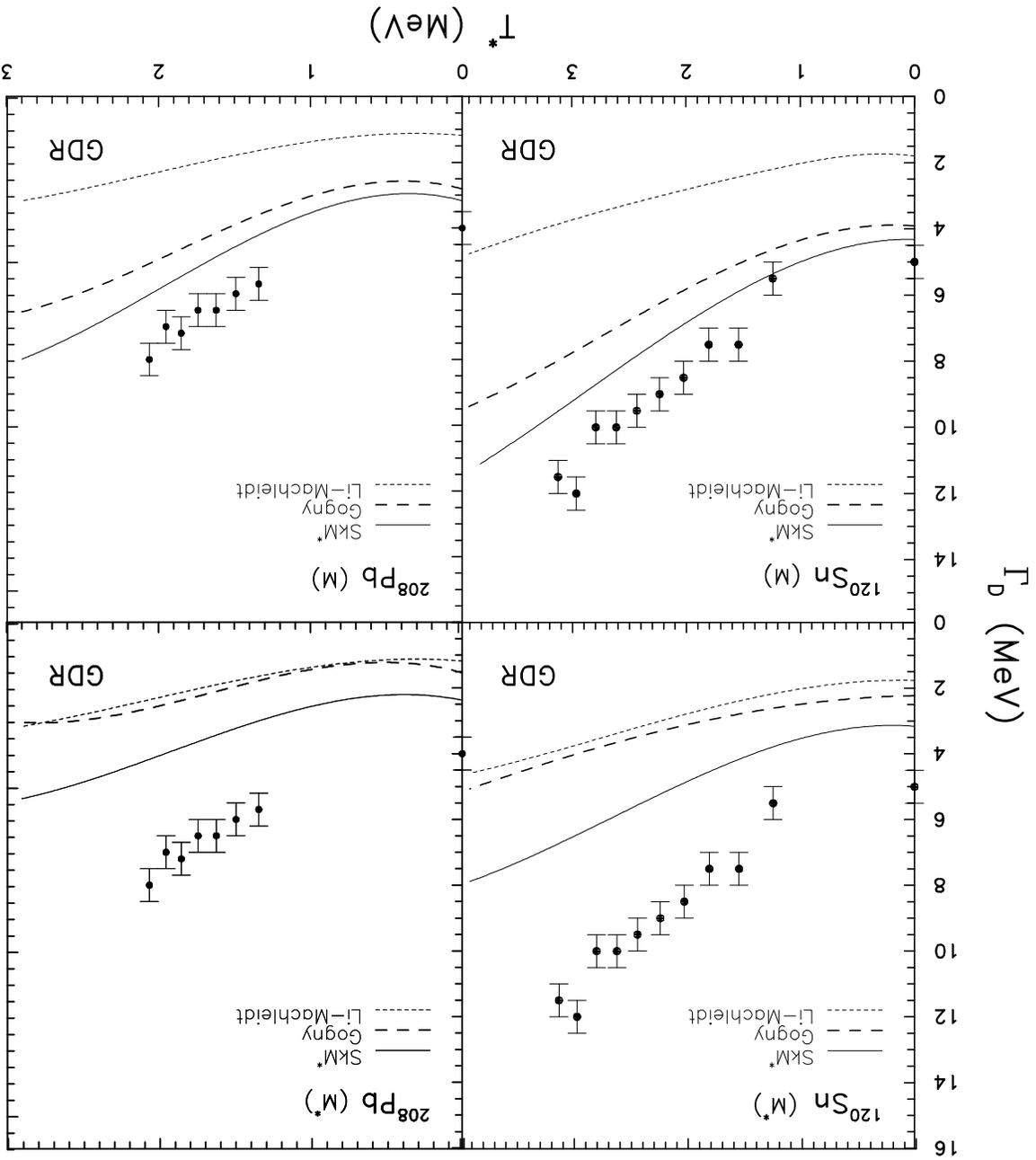
\item[{\bf Figure 3}:]  The collisional damping width of GDR in $^{120}Sn$
and $^{208}Pb$ as a function of temperature. Dotted lines are calculations
with the cross-sections of Li and Machleidt, and solid and dashed lines are
results with the SkM$^{*}$ and the Gogny cross-sections with the bare
nucleon mass (lower panel) and the effective nucleon mass (upper panel). The
data is taken from \cite{TinLead}.

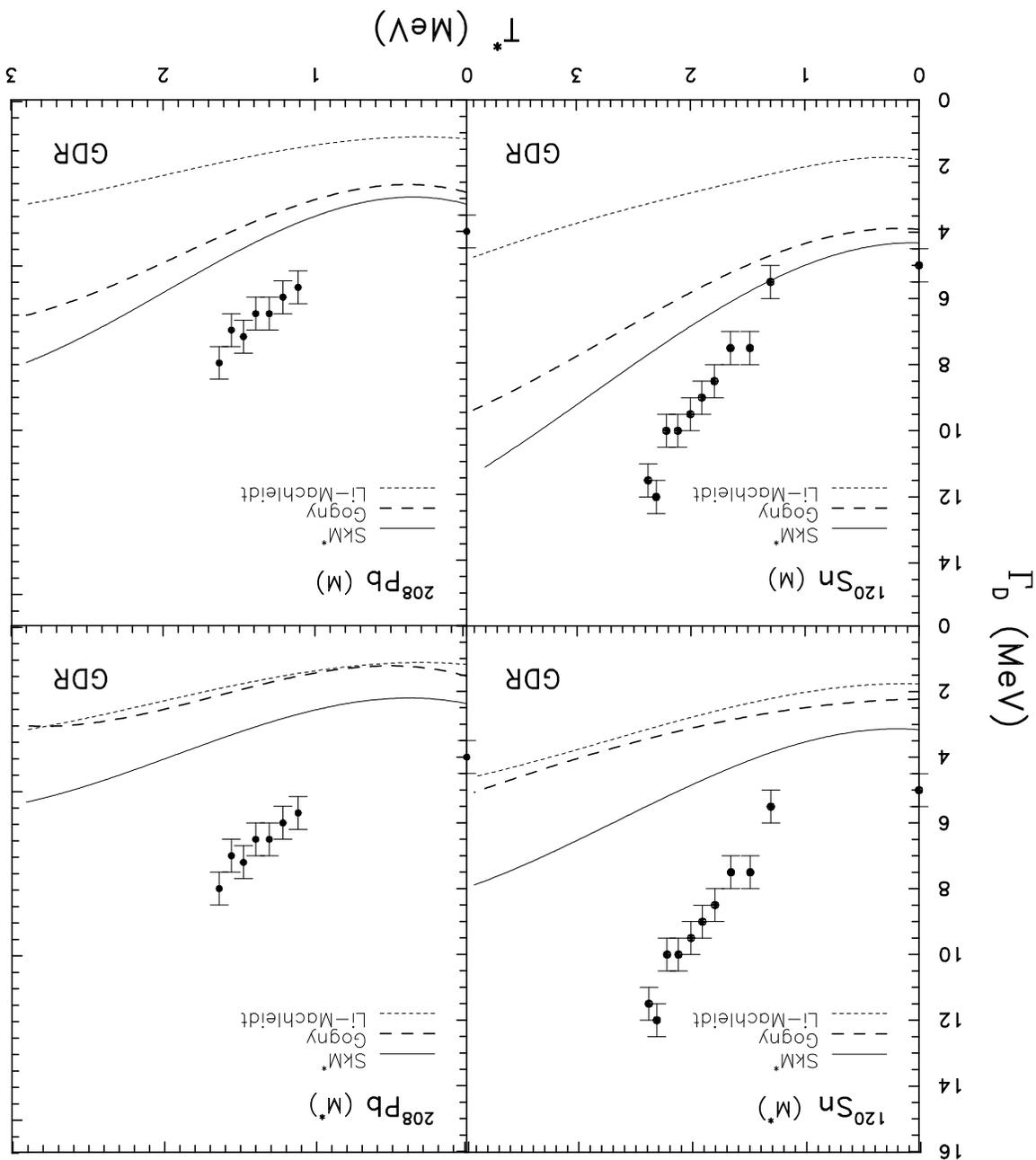
\item[{\bf Figure 4}:]  Same as in figure 3, but shown with the revised data
taken from \cite{Dimitri}.
\end{description}

\end{document}